\documentclass[twocolumn,floatfix,eqsecnum,aps,nofootinbib]{revtex4-2}
\usepackage{graphicx}
\usepackage{amsmath}
\usepackage{amssymb}
\usepackage{bm}
\usepackage{hyperref}

\usepackage{color}

\begin{document}

\title{Boundary-condition-assisted chiral-symmetry protection\\
 of the zeroth Landau level on a two-dimensional lattice}
\author{A. Don\'{i}s Vela}
\affiliation{Instituut-Lorentz, Universiteit Leiden, P.O. Box 9506, 2300 RA Leiden, The Netherlands}
\author{C. W. J. Beenakker}
\affiliation{Instituut-Lorentz, Universiteit Leiden, P.O. Box 9506, 2300 RA Leiden, The Netherlands}

\date{May 2025}

\begin{abstract}
The massless two-dimensional Dirac equation in a perpendicular magnetic field $B$ supports a $B$-independent ``zeroth Landau level'', a dispersionless zero-energy-mode protected by chiral symmetry. On a lattice the zero-mode becomes doubly degenerate with states of opposite chirality, which removes the protection and allows for a broadening when the magnetic field is non-uniform. It is known that this fundamental obstruction can be avoided by spatially separating the doubly degenerate states, adjoining $+B$ and $-B$ regions in a system of twice the size. Here we show that the same objective can be achieved without doubling the system size. The key ingredients are 1) a chirality-preserving ``tangent fermion'' discretization of the Dirac equation; and 2) a boundary condition that ensures the zero-mode contains only states of a single chirality.
\end{abstract}
\maketitle

\section{Introduction}

Attempts to discretize the massless Dirac equation
\begin{equation}
v(\bm{p}\cdot\bm{\sigma})\psi=E\psi \label{Diraceq}
\end{equation}
on a lattice face two fundamental obstructions: The first is the Nielsen-Ninomiya no-go theorem \cite{Nie81}, which states that any local discretization of the derivative operator $\bm{p}=-i\hbar\partial/\partial\bm{r}$ must either break chiral symmetry, or  introduce a spurious second branch of low-energy eigenstates (a lattice artefact known as fermion doubling \cite{Tong}). A nonlocal ``tangent fermion'' discretization can work around this first obstruction \cite{Bee23}.

The second obstruction is due to Stacey \cite{Sta83}: Any gauge-invariant, chiral-symmetry preserving lattice regularization of the Dirac equation in an unbounded system must have the same number of zero-energy-solutions (zero-modes) for either chirality. As a consequence, while in the continuum a magnetic field $B$ perpendicular to a two-dimensional (2D) plane introduces a zero-mode (zeroth Landau level) of a definite chirality \cite{Nie83}, on the lattice the zero-mode becomes doubly degenerate with states of opposite chirality. A topological index theorem \cite{Alv83} protects the zero-mode from broadening in a non-uniform magnetic field \cite{Kat08,Kai09}, but if both chiralities are present the topological index vanishes and the zero-mode is no longer protected.

In Ref.\ \onlinecite{Don23} a work-around for the second obstruction was proposed that doubles the size of the system by adjoining $+B$ and $-B$ regions, thereby spatially separating the states in the zeroth Landau level of opposite chirality. Locally a single chirality is present and the zero-mode is not broadened when $B$ has spatial fluctuations. 

In this work we develop an alternative means to preserve the chiral-symmetry protection of the zeroth Landau level, which does not require a doubling of the system size. That may be computationally advantageous, but more interestingly it works around Stacey's obstruction in a fundamentally different way.

Our approach is based on the observation that Stacey's no-go theorem assumes translational invariance \cite{Sta83}. Boundary conditions in a finite lattice \cite{Don25} may allow for a zero-mode of a definite chirality in zero magnetic field, so a nonzero topological index. The index theorem will then guarantee that the zero-mode persists as long as chiral symmetry is unbroken, hence we can expect an unbroadened zeroth Landau level in a non-uniform magnetic field.

In what  follows we demonstrate that this expectation works out. We apply the tangent fermion discretization of the Dirac equation to avoid fermion doubling, and impose the boundary condition $\psi=\sigma_z\psi$ that selects for a definite chirality. We then show by an explicit calculation that the zeroth Landau level remains unbroadened when the magnetic field varies from site to site on the lattice.

\section{Discretization of the Dirac equation on a finite lattice}

In this section we summarize results from Refs.\ \onlinecite{Don23} and \onlinecite{Don25} that we need in what follows.

\subsection{Gauge-invariant tangent fermions on an infinite lattice}

A gauge-invariant, chiral-symmetry-preserving discretization of the 2D Dirac equation was developed in Ref.\ \onlinecite{Don23}. Fermion doubling is avoided by a nonlocal Hamiltonian, with the tangent dispersion relation introduced by Stacey \cite{Sta82}. Computational efficiency is restored by a transformation of the nonlocal eigenvalue equation \eqref{Diraceq} into a local \textit{generalized} eigenvalue equation \cite{Pac21}, of the form ${\cal H}\psi=E{\cal P}\psi$ with local operators on both sides of the equation.

The operators ${\cal H}$ and ${\cal P}$ are given in terms of translation operators
\begin{equation}
{\cal T}_\alpha=\sum_{\bm n}e^{i\phi_\alpha(\bm{n})}|\bm{n}\rangle\langle \bm{n}+\bm{e}_\alpha|,\label{Talphadef}
\end{equation}
where the sum over $\bm{n}$ is a sum over lattice sites on a 2D square lattice and $\bm{e}_\alpha\in\{\bm{e}_x,\bm{e}_y\}$ is a unit vector in the $\alpha$-direction. The perpendicular magnetic field $B(\bm{r})\hat{z}=\nabla\times\bm{A}(\bm{r})$, with in-plane vector potential $\bm{A}=(A_x,A_y,0)$, enters via the phase
\begin{equation}
\phi_\alpha(\bm{n})=e\int_{\bm{n}+\bm{e}_\alpha}^{\bm{n}} A_\alpha(\bm{r})\,dx_\alpha.\label{phialphadef}
\end{equation}
We set the lattice constant $a$ and $\hbar$ to unity in most equations and denote the electron charge by $+e$. Notice that ${\cal T}_x$ and ${\cal T}_y$ do not commute,
\begin{equation}
{\cal T}_y{\cal T}_x=e^{2\pi i\varphi/\varphi_0}{\cal T}_x{\cal T}_y,
\end{equation}
where $\varphi$ is the flux through a unit cell in units of the flux quantum $\varphi_0=h/e$.

With these definitions the generalized eigenvalue equation is 
\begin{subequations}
\label{HPdefcalT}
\begin{align}
{\cal H}\psi={}&E{\cal P}\psi,\;\;{\cal P}={\cal C}{\cal C}^\dagger,\\
{\cal H}={}&\frac{\hbar v}{8ia}\sigma_x(1+{\cal T}_y)({\cal T}_x-{\cal T}_x^\dagger)(1+{\cal T}_y^\dagger)\nonumber\\
&+\frac{\hbar v}{8ia}\sigma_y(1+{\cal T}_x)({\cal T}_y-{\cal T}_y^\dagger)(1+{\cal T}_x^\dagger),\\
{\cal C}={}&\tfrac{1}{8}(1+{\cal T}_x)(1+{\cal T}_y)+\tfrac{1}{8}(1+{\cal T}_y)(1+{\cal T}_x).
\end{align}
\end{subequations}
The operator ${\cal H}$ is Hermitian and ${\cal P}={\cal C}{\cal C}^\dagger$ is Hermitian positive definite. These two properties ensure real eigenvalues $E_n$ and an orthonormal set of eigenstates
\begin{equation}
\Psi_n=Z_n{\cal C}^\dagger\psi_n,\;\;Z_n=\langle\psi_n|{\cal P}|\psi_n\rangle^{-1/2}.
\end{equation}

In zero magnetic field the tangent energy-momentum relation $E(k_x,k_y)$ follows from 
\begin{subequations}
\begin{align}
&\frac{\hbar v}{8ia}\sigma_x(1+e^{ik_y })(e^{ik_x} -e^{-ik_x })(1+e^{-ik_y})\psi\nonumber\\
&+\frac{\hbar v}{8ia}\sigma_y(1+e^{ik_x})(e^{ik_y } -e^{-ik_y })(1+e^{-ik_x })\psi\nonumber\\
&=E\left|\tfrac{1}{4}(1+e^{ik_x})(1+e^{ik_y })\right|^2\psi\\
&\Rightarrow E =\pm\frac{2\hbar v}{a}\sqrt{\tan^2(ak_x/2)+\tan^2(ak_y/2)}.
\end{align}
\end{subequations}
There is only a single Dirac cone, centered at $\bm{k}=0$, without spurious doublers at the Brillouin zone boundaries, while chiral symmetry is preserved (${\cal H}$ anticommutes with $\sigma_z$). There is no conflict with the Nielsen-Ninomiya no-go theorem \cite{Nie81}, since it does not apply to \textit{generalized} eigenvalue problems.

\subsection{Chirality-selective boundary condition in a finite lattice}

In Ref.\ \onlinecite{Don25} it was shown how boundary conditions on a domain $(x,y)\in{\cal D}$ can be introduced that preserve the Hermiticity of ${\cal H}$ and ${\cal P}$ as well as the positive definiteness of ${\cal P}$. Here we focus on the chirality-selective boundary condition
\begin{equation}
\psi(x,y)=\sigma_z\psi(x,y),\;\;(x,y)\in\delta D,\label{psiBC}
\end{equation}
which forces the spin-down component $\psi_\downarrow$ of the spinor $\psi=(\psi_\uparrow,\psi_\downarrow)$ to vanish on the boundary. In the context of graphene, Eq.\ \eqref{psiBC} characterizes the ``half-bearded graphene nanoribbon'' \cite{Wak01,Kus03,Koh07,Sar24}.

Any principal submatrix of a Hermitian positive definite matrix remains Hermitian positive definite. Starting from the infinite lattice, we remove the rows and columns of ${\cal H}$ and ${\cal P}$ that refer to sites outside of the system.\footnote{The order of operations is important here: We first apply the translation operators \eqref{Talphadef}  to the infinite lattice, and then restrict the ${\cal H}$ and ${\cal P}$ matrices to sites inside the system. This order ensures that ${\cal T}_\alpha$ is applied as a unitary operation, while it is not unitary on a finite lattice.} If the system contains $N$ sites the dimension of the resulting principal submatrices is $2N\times 2N$, including the spin degree of freedom. We denote by $N_B$ the number of sites on the boundary and by ${\cal B}$ the set of these lattice points.

 We implement the boundary condition \eqref{psiBC} on the lattice by removing from the matrices ${\cal H}$ and $ {\cal P}$ the spin-down row and column on each site $n\in{\cal B}$. The reduced matrices are
\begin{equation}
\tilde{\cal H}=Q^\top{\cal H}Q,\;\;\tilde{\cal P}=Q^\top{\cal P}Q,
\end{equation}
with $Q$ the rectangular matrix that projects out the $(2N-N_B)\times(2N-N_B)$ submatrix.

We thus arrive at the boundary-constrained generalized eigenvalue problem
\begin{equation}
\tilde{\cal H}{\psi}=E\tilde{\cal P}{\psi}.\label{tildeHP}
\end{equation}
The orthonormal eigenstates are
\begin{equation}
{\Psi}_n={Z}'_n{\cal C}^\dagger Q\psi_n,\;\;{Z}'_n=\langle\psi_n|\tilde{\cal P}|\psi_n\rangle^{-1/2}.
\end{equation}

\begin{figure}[tb]
\centerline{\includegraphics[width=0.8\linewidth]{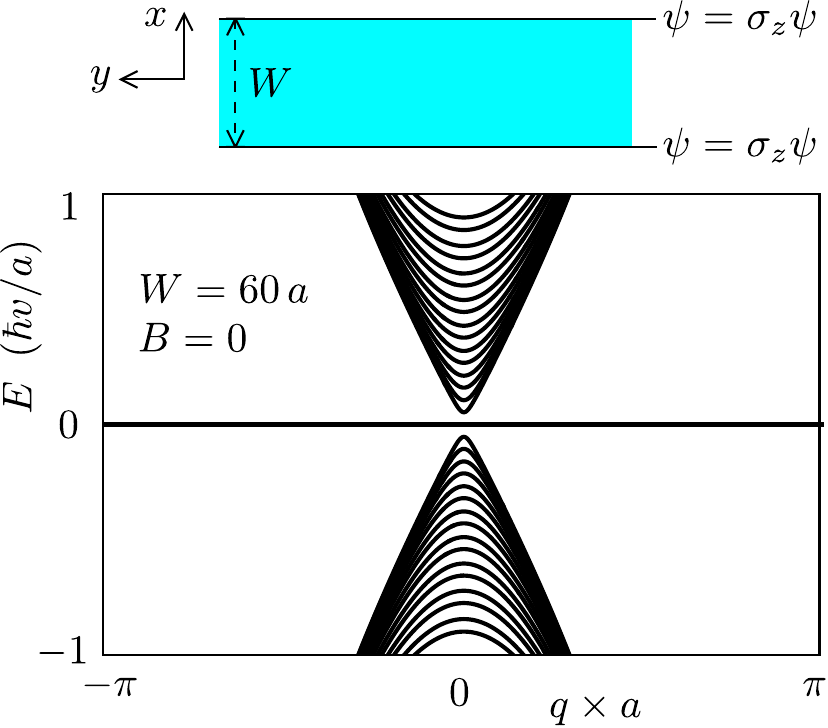}}
\caption{Dispersion relation on a lattice confined to an infinitely long channel of 60 lattice points across, with boundary condition $\psi=\sigma_z\psi$ on both edges. The energy is plotted as a function of the momentum $q=k_y$ along the channel. This is for zero magnetic field, the zero-energy mode then consists of a pair of spin-up edge modes.
}
\label{fig_zerofield}
\end{figure}

\begin{figure}[tb]
\centerline{\includegraphics[width=0.8\linewidth]{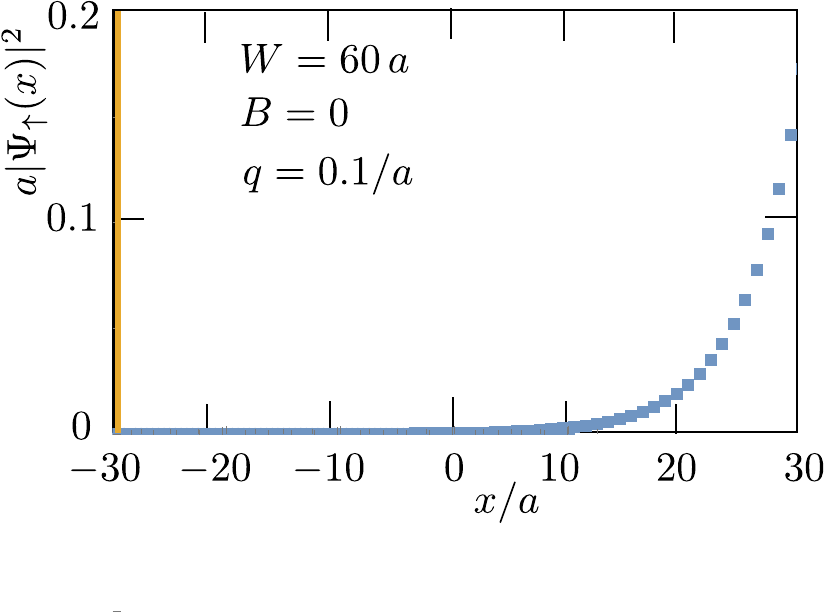}}
\caption{Spin-up polarized wave function profile $|\Psi_\uparrow|^2$ in the zero-mode of Fig.\ \ref{fig_zerofield}, at momentum $q=0.1/a$. The edge mode at $x=\operatorname{sign}(q)W/2$ of width $ 1/q$ (blue markers) transforms into a zeroth Landau level upon application of a magnetic field. The edge mode at the opposite edge (orange), just one lattice constant wide, is a lattice artefact. It does not contribute to the zeroth Landau level. 
}
\label{fig_profile}
\end{figure}

In Fig.\ \ref{fig_zerofield} we show the resulting band structure in a channel geometry, with boundaries at $x=\pm W/2$. The parallel momentum $k_y\equiv q$ is a good quantum number, so to compute the dispersion relation $E(q)$ we can work on a 1D lattice (sites $n=1,2,\ldots N\equiv W/a$, Brillouin zone $-\pi/a<q<\pi/a$). A doubly degenerate zero-energy mode appears, see Fig.\ \ref{fig_profile}, localized on the edges of the channel and fully spin-up polarized \cite{Wak01,Kus03,Koh07,Sar24}. As we will now show, upon application of a magnetic field one of the two edge modes moves to the interior of the channel, forming a spin-polarized zeroth Landau level.

\section{Polarized zeroth Landau level}

\subsection{Zeroth Landau level in the continuum}

We recall textbook results for the Landau level spectrum of massless Dirac fermions \cite{Kat12}. In an infinite 2D system, the eigenvalues of the Dirac Hamiltonian
\begin{equation}
{ H}_{\rm Dirac}=v(\bm{p}-e\bm{A})\cdot\bm{\sigma}
\end{equation}
are dispersionless flat bands at energies
\begin{equation}
E_n=\pm v\sqrt{2n\hbar e|B|},\;\;n=0,1,2,\ldots.\label{Endef}
\end{equation}
Both chiralities contribute equally to each nonzero Landau level: $\langle n|\sigma_z|n\rangle=0$ for $n\geq 1$. The zeroth Landau level, however, is polarized: $\langle 0|\sigma_z|0\rangle={\rm sign}\,(B).$

This property of the zeroth Landau level follows from a simple calculation. Rewrite the Dirac equation ${ H}_{\rm Dirac}\psi=E\psi$, in the gauge $\bm{A}=Bx\hat{y}$, as an ordinary differential equation,
\begin{equation}
\frac{d}{dx}\psi(x)=i\sigma_x[E/v-(q-eBx)\sigma_y]\psi(x),
\end{equation}
which for $E=0$ has the solution
\begin{equation}
\psi(x)=\exp\left(-\tfrac{1}{2}eB(x-q/eB)^2\sigma_z\right)\psi(q/eB).\label{psixB}
\end{equation}
For a normalizable solution we need $\sigma_z\psi=\operatorname{sign}(B)\psi$, so the zeroth Landau level is spin polarized. At a given $q$ the state is localized along the line $x=q/eB$, with decay length $l_m=\sqrt{\hbar/e|B|}$ (the magnetic length).

\subsection{Zeroth Landau level on a lattice}

The polarization of the zeroth Landau level is lost on an infinite lattice, instead at each $q$ there are two zero-energy states of opposite chirality. A topological argument for this goes as follows \cite{Don25}. 

The difference ${\cal I}(B)$ of the number of spin-up and spin-down states in the zero-mode is a topological invariant. If we smoothly deform the uniform magnetic field $B$, to make it non-uniform, ${\cal I}$ cannot change. Now imagine concentrating the magnetic flux lines, in such a way that a unit cell of the lattice either encloses zero flux or one flux quantum. The field can then be removed by a gauge transformation, so we conclude that, on the lattice
\begin{equation}
{\cal I}(B)={\cal I}(0).
\end{equation}
Since ${\cal I}(0)=0$ in the infinite lattice, the zeroth Landau level must contain an equal number of spin-up and spin-down states at any nonzero $B$.

This argument directly points to a way out in a finite lattice: If the lattice has a polarized zero-mode at $B=0$, the zeroth Landau level that forms in a field will be polarized to ensure that the topological invariant remains unchanged.

\begin{figure}[tb]
\centerline{\includegraphics[width=0.8\linewidth]{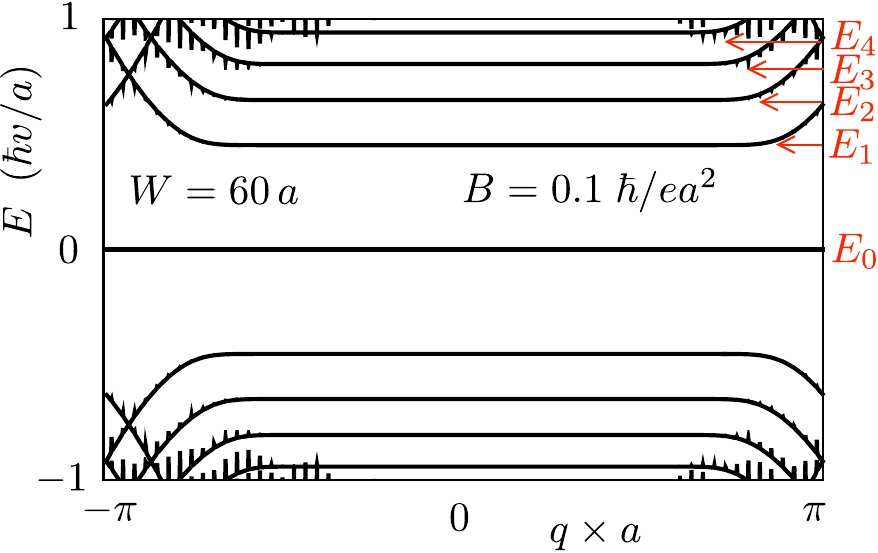}}
\caption{Same as Fig.\ \ref{fig_zerofield}, but now in a perpendicular magnetic field corresponding to $0.1/2\pi$ flux quanta $h/e$ through each unit cell of the lattice. Dispersionless Landau levels form at low energies (the sharp peaks at high energies are a lattice artefact). Red arrows indicate the location of the Landau levels in the continuum, according to Eq.\ \eqref{Endef}. All levels are non-degenerate, except the zero-energy mode, which is doubly degenerate.
}
\label{fig_nonzerofield}
\end{figure}

\begin{figure}[tb]
\centerline{\includegraphics[width=0.8\linewidth]{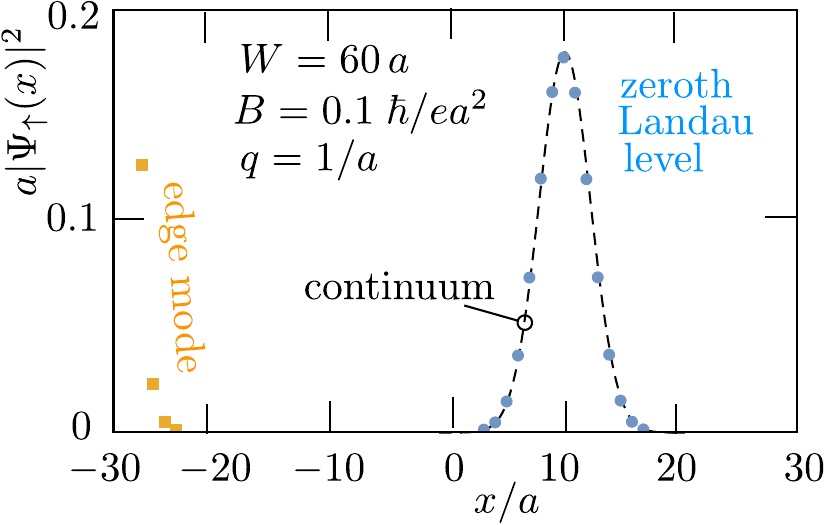}}
\caption{Spin-up polarized wave function profile $|\Psi_\uparrow|^2$ in the zero-mode of Fig.\ \ref{fig_nonzerofield}, at momentum $q=1/a$. This doubly degenerate level contains an edge mode (square markers) and a mode centered at $x=q\hbar/eB=10\,a$ (round markers). The dashed curve is the profile of the zeroth Landau level in the continuum, according to Eq.\ \eqref{psixB}.
}
\label{fig_profileB}
\end{figure}

We demonstrate it in Figs.\ \ref{fig_nonzerofield} and \ref{fig_profileB}. The zero-mode at $B=0$ has ${\cal I}=2$, consisting of two spin-up polarized edge modes. Upon application of a magnetic field one of these edge modes moves into the interior of the channel, towards the $q$-dependent position at $x=q/eB$. This is the expected result for a zeroth Landau level in the continuum, see Eq.\ \eqref{psixB}.  

The magnetic field should be neither too small nor too large for this to work. On the one hand, we need $a\ll l_m\ll W$ to minimize lattice and boundary effects. On the other hand, we need $BWa<h/e$ to avoid a folding of the Brillouin zone:\footnote{The condition $BWa<h/e$ is for an infinitely long channel. More generally, in a 2D geometry of surface area $S$ and perimeter $P$ (with boundary condition $\psi=\sigma_z\psi$ along the entire perimeter), the upper limit of the field strength follows by noting that the zeroth Landau  level of degeneracy $BSe/h$ is formed out of an edge mode of degeneracy $P/a$, hence $BSe/h$ cannot be larger than $P/a \Rightarrow B<(h/e)P/Sa$.
}
 the appearance of new zero-energy states at $x= (q\pm 2\pi/a)(\hbar/eB)$. For $W\gg a$ this leaves a broad range
\begin{equation}
a\ll l_m < (Wa/2\pi)^{1/2}.
\end{equation}
of applicability.

\section{Disorder effects}

\subsection{Spatially varying magnetic field}

The nonzero topological invariant ${\cal I}=2$ should protect the zero-mode from broadening in the presence of disorder that preserves chiral symmetry. To test this, we consider a spatially fluctuating magnetic field, varying uniformly from site to site in the interval $(B_0-\delta B,B_0+\delta B)$. The corresponding vector potential is implemented on the lattice by adding a random flux $\delta\varphi\in(-a^2\delta B,a^2\delta B)$ through each unit cell, and then computing the corresponding phase shifts \eqref{phialphadef}. We show data for a single disorder realization (no disorder averaging).

\begin{figure}[tb]
\centerline{\includegraphics[width=0.8\linewidth]{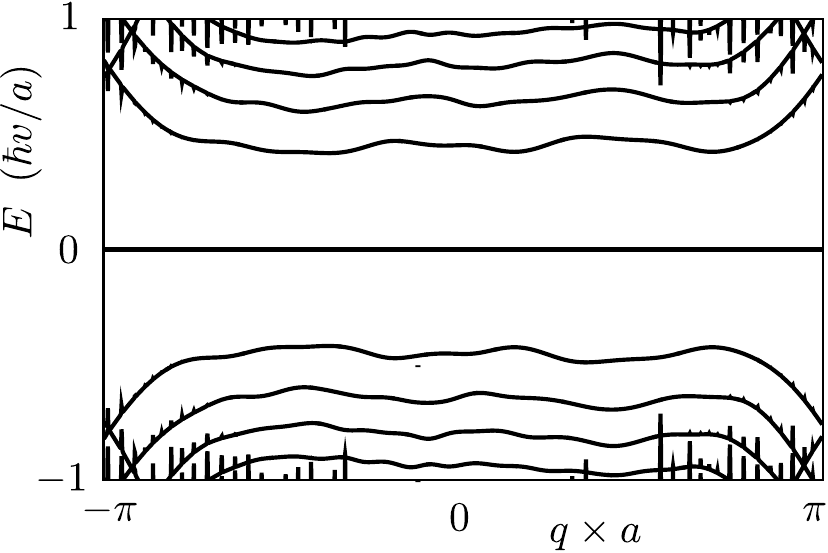}}
\caption{Same as Fig.\ \ref{fig_nonzerofield}, but now for a magnetic field that varies randomly across the channel, $B(x)\in(B_0-\delta B,B_0+\delta B)$, with $B_0=0.1\,\hbar/ea^2=2\,\delta B$.
}
\label{fig_disorder}
\end{figure}

We first keep the field $B(x)$ translationally invariant along the channel, so that $q$ remains a good quantum number and we can work with the dispersion relation $E(q)$. Fig.\ \ref{fig_disorder} shows the Landau levels for $\delta B=\tfrac{1}{2}B_0$. The Landau levels at nonzero energy are significantly broadened. The zero-mode, instead, remains precisely flat. We have checked that there is also no significant effect on the wave function profile, we retain the zeroth Landau level peak centered at $x=q\hbar/eB_0$.

\begin{figure}[tb]
\centerline{\includegraphics[width=0.8\linewidth]{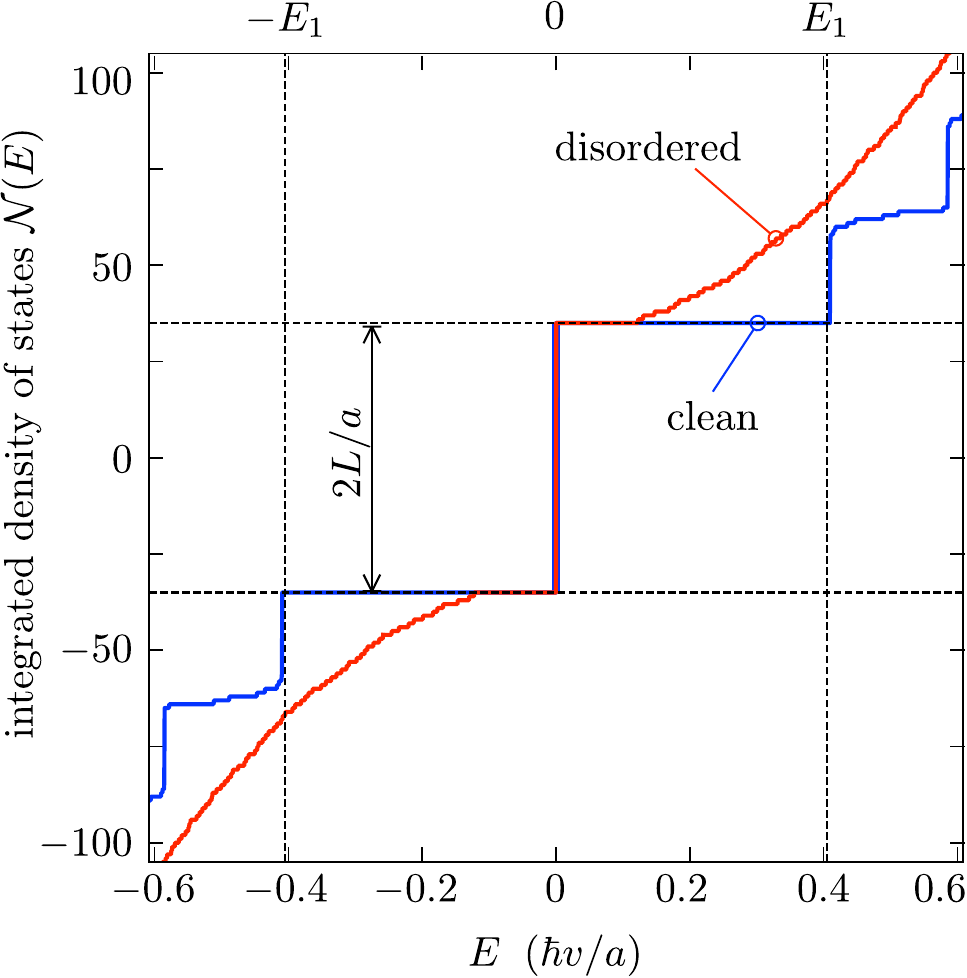}}
\caption{Integrated density of states in a rectangular geometry, $W\times L=65\,a\times 35\,a$, with boundary condition $\psi=\sigma_z\psi$ at $x=\pm W/2$ and periodic boundary conditions at $y=\pm L/2$. In the clean system the magnetic field is uniform $B_0=0.85\times(2\pi\hbar/eaW)$; in the disordered system the field varies randomly from site to site, uniformly in the interval $(B_0-\delta B,B_0+\delta B)$ with $\delta B=10\,B_0$. The vertical dashed lines indicate the energy \eqref{Endef} of the first Landau level. The horizontal dashed lines mark the degeneracy $2L/a$ of the zero-energy mode. At this magnetic field value the degeneracy of the zeroth Landau level is $0.85\,L/a$ (enclosed flux $\Phi=BLW$ divided by $\varphi_0=h/e$). The additional $1.15 \,L/a$ states in the zero-mode are edge states. All $2L/a$ zero-energy states have a single chirality, so the zero-mode is not broadened by disorder.
}
\label{fig_DOS_tangent}
\end{figure}

We next let $B(x,y)$ vary in all directions, in a rectangular geometry $|x|<W/2$, $|y|<L/2$. In the $x$-direction we keep the chirality-selective boundary condition $\psi(\pm W/2,y)=\sigma_z\psi(\pm W/2,y)$, in the $y$-direction we impose periodic boundary conditions, $\psi(x,L/2)=\psi(x,-L/2)$. 

In Fig.\ \ref{fig_DOS_tangent} we plot the integrated density of states
\begin{equation}
{\cal N}(E)=\int_0^E \rho(E')\,dE',
\end{equation}
which in the clean system shows a step when $E$ crosses the Landau level energy $E_n$. The zero-energy mode remains completely flat in the presence of a randomly fluctuating magnetic field --- even at the relatively large disorder strength $\delta B=10\,B_0$.

\subsection{Comparison with graphene}

\begin{figure}[tb]
\centerline{\includegraphics[width=0.8\linewidth]{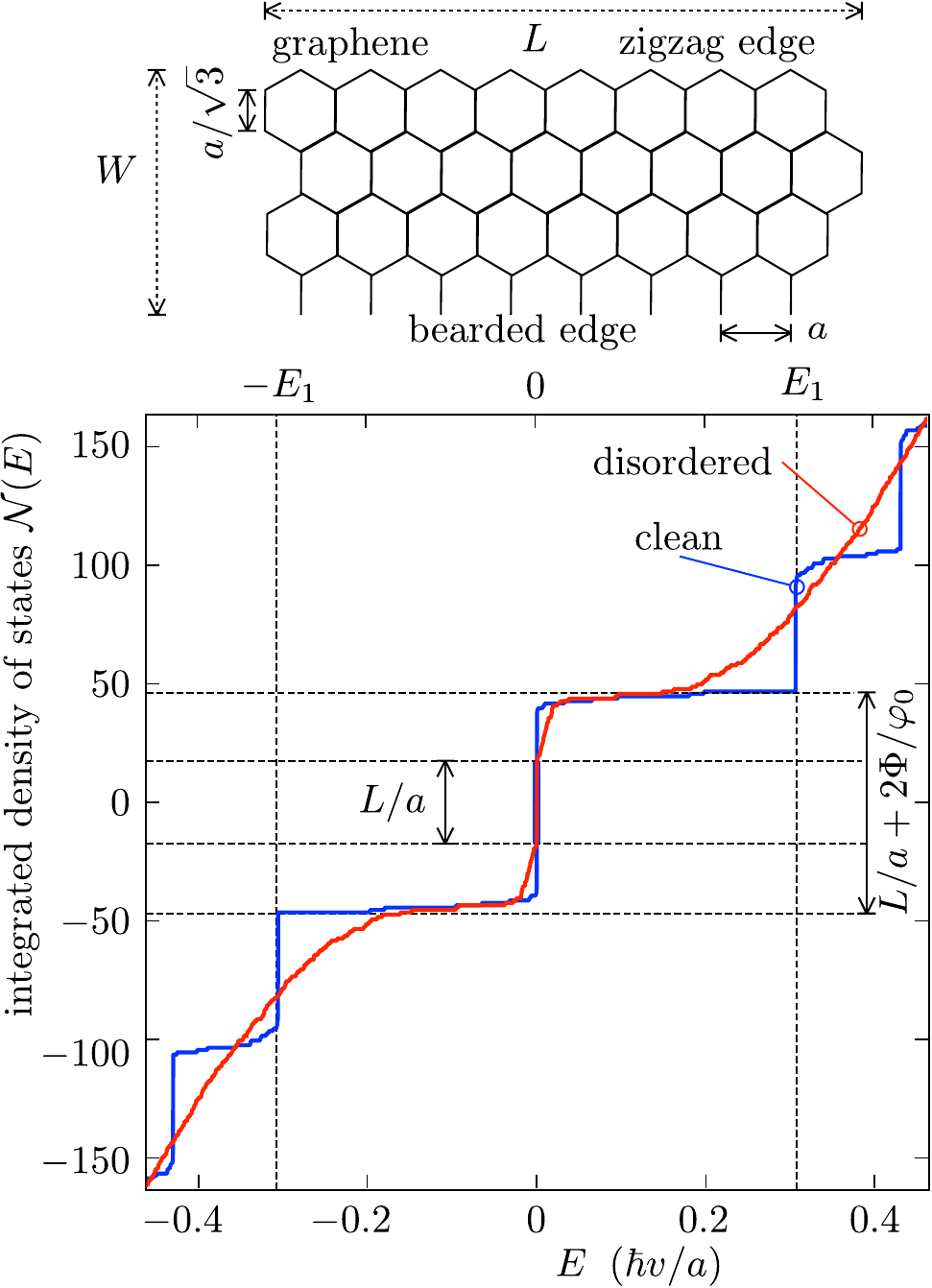}}
\caption{Same as Fig.\ \ref{fig_DOS_tangent}, but for Dirac fermions on the honeycomb lattice of graphene, rather than for tangent fermions on a square lattice. The combination of a zigzag edge at $x=W/2$ and a bearded edge at $x=-W/2$ ensures that the boundary condition is $\psi=\sigma_z\psi$ on both edges. Along $y$ we have again periodic boundary conditions. The dimensions of the channel are $W\times L=65\sqrt{3}\,a\times 35\,a$ (the width is $\sqrt 3$ larger than in Fig.\ \ref{fig_DOS_tangent} to accomodate an integer number of unit cells). The uniform magnetic field in the clean system is $B_0=0.85\times (2\pi \hbar/eaW)$, the disorder strength is again $\delta B=10\,B_0$. The edge mode has degeneracy $L/a$, it has a single chirality and is protected by chiral symmetry. The zeroth Landau level, in contrast, combines states of both chiralities and therefore lacks protection from disorder broadening.
}
\label{fig_DOS_graphene}
\end{figure}

In graphene the geometry of Fig.\ \ref{fig_DOS_tangent} is a nanoribbon, closed into a ring of length $L$, with a bearded edge at $x=-W/2$ and a zigzag edge at $x=W/2$. The zeroth Landau level in graphene is not topologically protected because of the presence of two Dirac cones in the Brillouin zone \cite{Wat10,Kaw10,Per11,Kaw13}. The pair of Dirac cones each contribute a Landau level of opposite chirality, with net zero contribution to the topological index. In Fig.\ \ref{fig_DOS_graphene} we check that, indeed, a spatially varying magnetic field broadens the zeroth Landau level. Only the spin-polarized edge state remains flat.

\section{Conclusion}

In conclusion, we have presented a method to discretize the Dirac equation on a lattice which a) preserves chiral symmetry and b) supports a zeroth Landau level of a definite chirality. By working with a finite system, we circumvent Stacey's obstruction \cite{Sta83}, that a zero-mode on an infinite lattice cannot have a definite chirality. The combination of a) and b) protects the zeroth Landau level from broadening in the presence of a spatially varying magnetic field, in view of a topological index theorem \cite{Alv83,Kat08,Kai09}.

In a three-dimensional topological insulator the zero-modes of opposite chirality are spatially separated on opposite surfaces \cite{Yan11,Bre14,Xu14,Kon14}, which is an altogether different way to work around the fermion doubling obstruction. Our strictly two-dimensional approach offers a computationally less expensive work around, that may also be useful for studies of interacting chiral fermions \cite{Zak24a,Hae24,Zak24b}.

\acknowledgments

Discussions with G. Lemut and M. J. Pacholski are gratefully acknowledged. This project was supported by the Netherlands Organisation for Scientific Research (NWO/OCW), as part of Quantum Limits (project number SUMMIT.1.1016).

The computer codes used for Figs.\ \eqref{fig_DOS_tangent} and \eqref{fig_DOS_graphene} are available at \url{https://dx.doi.org/10.5281/zenodo.15456629}.

\end{document}